\documentclass[twocolumn,reprint,superscriptaddress,amsmath,amssymb,aps,prb]{revtex4-2}

\usepackage{dcolumn}
\usepackage{bm}
\usepackage[T1]{fontenc}
\usepackage[english]{babel}
\usepackage[utf8]{inputenc}
\usepackage{longtable}
\usepackage{xspace,graphicx}
\newcommand{\invcm}{cm\ensuremath{^{-1}}\xspace}

\newcommand{\NH}{\hbox{NH$_3$}\xspace}
\newcommand{\NHP}{\hbox{NH$_4^+$}\xspace}

\newcommand{\HHO}{\hbox{H$_2$O}\xspace}
\newcommand{\OHm}{\hbox{OH$^-$}\xspace}
\newcommand{\Odm}{\hbox{O$^{2-}$}\xspace}

\newcommand{\micron}{\ensuremath{\mu}m\xspace}
\newcommand{\bcc}{\textit{bcc}\xspace}
\graphicspath{{./fig/}}
\addto\captionsenglish{}
\usepackage[squaren, Gray, cdot]{SIunits}
\usepackage{lipsum}
\usepackage{xcolor}
\usepackage[%
  colorlinks=true,
  urlcolor=blue,
  linkcolor=blue,
  citecolor=blue
]{hyperref}
\usepackage{comment}

\newcommand{\pbcc}{\emph{p}bcc\xspace}
\newcommand{\qbcc}{\emph{q}bcc\xspace}

\begin{document}
\preprint{APS/123-QED}

\title{High pressure-temperature phase diagram of ammonia hemihydrate}

\author{L. Andriambariarijaona}
\affiliation{Institut de Min\'{e}ralogie, de Physique des Mat\'{e}riaux et de Cosmochimie (IMPMC), Sorbonne Universit\'{e}, CNRS UMR 7590,  MNHN, 4, place Jussieu, Paris, France}
\author{F. Datchi}
\affiliation{Institut de Min\'{e}ralogie, de Physique des Mat\'{e}riaux et de Cosmochimie (IMPMC), Sorbonne Universit\'{e}, CNRS UMR 7590,  MNHN, 4, place Jussieu, Paris, France}
\author{H. Zhang}
\affiliation{Institut de Min\'{e}ralogie, de Physique des Mat\'{e}riaux et de Cosmochimie (IMPMC), Sorbonne Universit\'{e}, CNRS UMR 7590,  MNHN, 4, place Jussieu, Paris, France}
%\affiliation{State Key Laboratory of Superhard Materials, Jilin University, Changchun 130012, China}
\author{K. B\'{e}neut}
\affiliation{Institut de Min\'{e}ralogie, de Physique des Mat\'{e}riaux et de Cosmochimie (IMPMC), Sorbonne Universit\'{e}, CNRS UMR 7590, MNHN, 4, place Jussieu, Paris, France}
\author{B. Baptiste}
\affiliation{Institut de Min\'{e}ralogie, de Physique des Mat\'{e}riaux et de Cosmochimie (IMPMC), Sorbonne Universit\'{e}, CNRS UMR 7590, MNHN, 4, place Jussieu, Paris, France}
\author{N. Guignot}
\affiliation{Synchrotron SOLEIL, BP 48, 91192 Gif Sur Yvette, France}
\author{S. Ninet}
\affiliation{Institut de Min\'{e}ralogie, de Physique des Mat\'{e}riaux et de Cosmochimie (IMPMC), Sorbonne Universit\'{e}, CNRS UMR 7590,  MNHN, 4, place Jussieu, Paris, France}

\date{\today}

\begin{abstract}
We report a comprehensive experimental investigation of the phase diagram of ammonia hemihydrate (AHH) in the range of 2-30 GPa and 300-700 K, based on Raman spectroscopy and x-ray diffraction experiments and visual observations. Four solid phases, denoted AHH-II, DIMA, \pbcc and \qbcc, are present in this domain, one of which, AHH-\qbcc was discovered in this work. We show that, unlike previously thought, the body-centered cubic (bcc) phase obtained on heating AHH-II below 10 GPa, denoted here as AHH-\pbcc, is distinct from the DIMA phase, although both present the same bcc structure and O/N positional disorder. Our results actually indicates that AHH-\pbcc is a plastic form of DIMA, characterized by free molecular rotations.  AHH-\qbcc is observed in the intermediate P-T range between AHH-II and DIMA. It presents a complex x-ray pattern reminiscent of the ''quasi-bcc'' structures that have been theoretically predicted, although none of these structures is consistent with our data. The transition lines between all solid phases as well as the melting curve have been mapped in detail, showing that: (1)  the new \qbcc phase is the stable one in the intermediate P-T range 10-19 GPa, 300-450 K, although the II-\qbcc transition is kinetically hindered for $T<450$ K, and II directly transits to DIMA in a gradual fashion from 25 to 35 GPa at 300 K. (2) The stability domain of \qbcc shrinks above 450 K and eventually terminates at a \pbcc-qbcc-DIMA triple point at 21.5 GPa-630 K.  (3) A direct and reversible transition occurs between AHH-\pbcc and DIMA above 630  K. (4) The \pbcc solid stability domain extends up to the melting line above 3 GPa, and a II--\pbcc--liquid triple point is identified at 3 GPa-320 K. 
\end{abstract}

\maketitle
\section{Introduction}
Water and ammonia are two polar molecules that mix in any proportion in the liquid state. Upon cooling the liquid mixtures at 1 bar, three stable stoichiometric compounds may crystallize depending on the concentration: ammonia monohydrate (AMH, \HHO:\NH), ammonia hemihydrate (AHH, \HHO:2\NH), and ammonia dihydrate (ADH, \HHO:2\NH). These compounds have attracted a great interest due to the fact that they are prototypical hydrogen-bonded solids exhibiting heteronuclear O-H...N and N-H..O bonds, on one hand, and because they are potential mineral phases of icy planets and satellites of our solar system, on the other hand. Indeed, planets like Uranus and Neptune, or satellites like Titan and Ganymede are believed to host enormous amounts of water, ammonia and methane in their lower mantle, some of which might be in solid form. The relative abundance of water to ammonia in the solar system, about 7\HHO:1\NH, explains that the majority of previous studies  have focused on the water-rich ammonia hydrates, i.e. ADH and AMH~\cite{Lunine1987,Boone,LovedayPRL_1999,Fortes2007,Loveday2009,Liu2017,Zhang2020,Zhang2023}.

However, experiments have shown that both AMH and ADH decompose into AHH + ice VII upon compression of the liquid at ambient T and low pressure (P$<$3 GPa) \cite{Boone, Fortes2007, Wilson2012, Zhang2020}. Moreover, numerical simulations have suggested that the ammonia-rich AHH solid is more stable than AMH and ADH at the extreme P-T conditions encountered in giant planets \cite{Naden2017}. This could imply that the abundance of AHH within planetary interiors is larger than previously thought.

\begin{figure}
\centering
\includegraphics[width=1\linewidth]{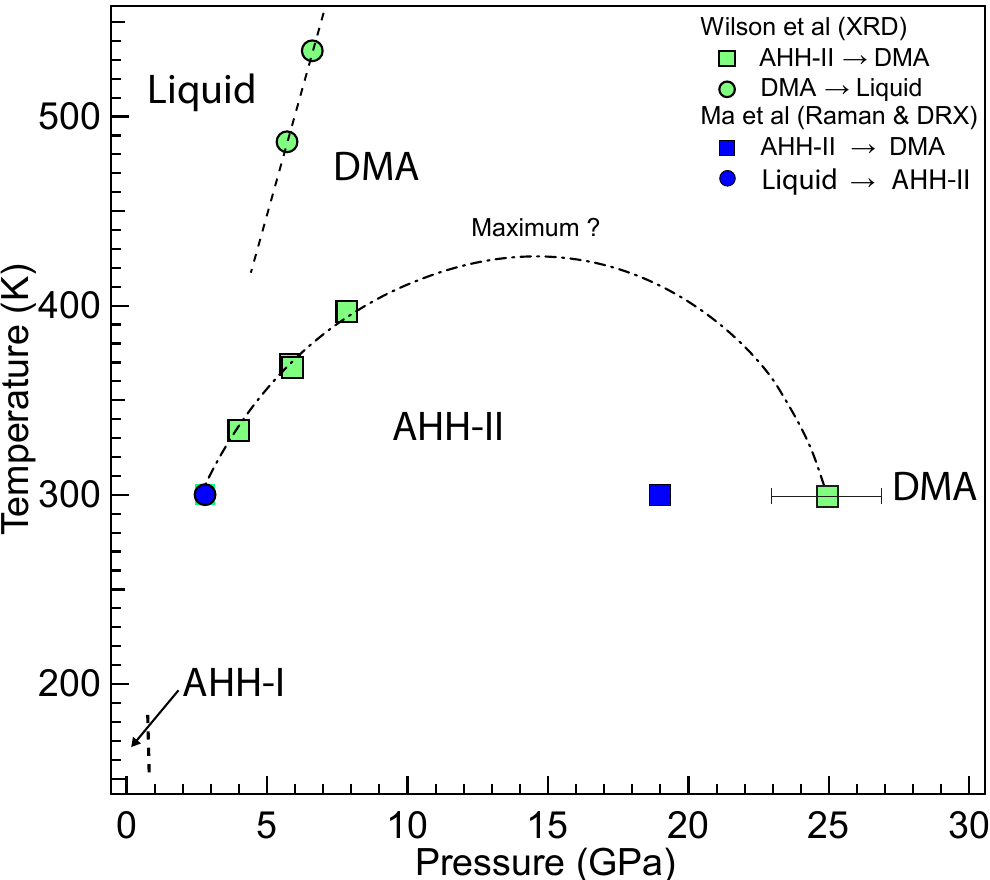}
\caption{\label{fig:AHH_Diagram_Intro} Experimental phase diagram of AHH as known at the beginning of this study, adapted from Ref.\cite{Wilson2015,ma2012}. The dashed-dotted line represents the AHH-II DMA transition line suggested by \citet{Wilson2015}. }
\end{figure}

The current knowledge on the high pressure phase diagram of AHH is feeble and mostly limited to room temperature. As shown in Fig.~\ref{fig:AHH_Diagram_Intro}, four solid phases, named AHH-I , AHH-II , DMA (Disordered Molecular Alloy), and AHH-III  have been reported. AHH-I is an orthorhombic molecular crystal stable between 1 bar and 1 GPa and temperatures below 200 K \cite{SiemonsAC_1954,LovedayHPR_2003}. AHH-II is obtained by compressing the liquid at 300 K and is a molecular crystal with a monoclinic space group P2$_1$/c and 12 molecules per unit cell \citep{Wilson2012,Wilson2015}. According to \citet{Wilson2015}, AHH-DMA is obtained on compressing AHH-II at 300 K above 25(2) GPa. This phase, also observed in the 1:1 (AMH) and 1:2 (ADH) solid has been initially described as a disordered molecular alloy (DMA) constructed by randomly positioning water and ammonia molecules on a body-centered cubic (bcc) lattice  \cite{LovedayPRL_1999}. \citet{Liu2017} later showed that the DMA phase in AMH is partially ionic and composed of one-third OH$^-$ and NH$^+_4$ ions and two-third water and ammonia molecules, so DMA was rebaptized DIMA for Disordered Ionico-Molecular Alloy. More recently, \citet{XuPRL2021} found evidence that the AHH-DMA is also partially ionic, so we will refer to this phase as AHH-DIMA throughout the manuscript. Interestingly, \citet{Wilson2015} also found that AHH-II transits into a bcc solid upon heating in the range 4-8 GPa, with the transition temperature rising from 350 K to 400 K. Assimilating this bcc solid to AHH-DIMA, they postulated that the AHH-II to DIMA transition temperature initially rises with pressure, goes through a maximum and then downturns to reach 300 K around 25 GPa. However, the transition line has so far not been investigated beyond 8 GPa. Finally, another phase called AHH-III, has been reported at 300~K above 69 GPa~\cite{XuPRL2021}. Based on Raman spectroscopy and first-principles calculations, \citet{XuPRL2021} assigned AHH-III to the predicted $P\bar{3}m1$ ionic  phase\cite{Naden2017} with formula (\NHP)$_2$\Odm , which has not been confirmed by other groups to date.

In this work, we performed a comprehensive set of complementary experiments, namely x-ray diffraction (XRD), Raman spectroscopy, and visual observations, that allowed us to obtain the first complete experimental phase diagram of AHH in the P-T domain 2--30 GPa, 300--700 K. Four solid phases, denoted AHH-II, AHH-DIMA, AHH-\pbcc and AHH-\qbcc, are observed and characterized by XRD and Raman spectroscopy. The transition lines between all solid phases, as well as the solid-liquid (melting) line are precisely determined. AHH-\qbcc was discovered in the course of this work and was found to be the stable phase in the intermediate P-T range between AHH-II and DIMA. The structure of AHH-\qbcc is complex and doesn't correspond to any of the AHH structures predicted stable in this pressure range \cite{Naden2017}. We also demonstrate that the two \bcc, DIMA-like, phases obtained respectively by compression and by heating of phase II and assumed to be the same phase are in fact two  distinct ones, AHH-DIMA and AHH-\pbcc. Above 630~K and 21~GPa, an isostructural phase transition is observed between AHH-DIMA and AHH-\pbcc. Finally, several arguments strongly suggest that the AHH-\pbcc corresponds to a plastic form of AHH-DIMA.

This paper is organized as follows. Section II presents the experimental methods. In section III, we describe our experimental investigations of the solid-solid and solid-liquid transition lines and the evidence for the new AHH-\qbcc phase. In section IV, we discuss the plastic nature of AHH-\pbcc and the structural and vibrational information that have been obtained on AHH-\qbcc.

\section{Experimental methods}

To generate high pressure conditions, we used membrane diamond anvil cells (mDAC) equipped with diamond anvils of 0.3 to 0.6 mm culet diameters, and 0.2 mm-thick Re foils as gaskets. Gaskets were pre-indented and drilled with a hole of 0.1 to 0.2 mm to serve as sample chamber. The latter was lined with gold to prevent any possible chemical reaction of ammonia or water with rhenium at high temperature. The gold liner was also used as pressure calibrant during XRD experiments.

(1\HHO, 2\NH) molar mixtures were prepared and cryoloaded in the mDAC with the same methods as described in Refs.~\cite{Liu2017,Zhang2020}. The accuracy on the molar concentration of the mixture is better than 1\%. The mixture is cooled to 210 K before a drop of liquid is poured into the gasket hole while the mDAC is maintained at about 120 K. The mDAC is then rapidly closed and a sufficient load is applied at low T to ensure that the sample is sealed before warming up to 300 K. 

The pressure was determined with the luminescence of ruby spheres using the Ruby2020 calibration \cite{Shen_2020}, or with the thermal equation of state of gold from Ref.\cite{Dorogokupets2007}. The mDAC was inserted inside a commercial cylindric resistive heater to heat the sample up to 700 K. The temperature of the sample was measured with a chromel-alumel thermocouple positioned inside the mDAC next to the diamond anvils.

Raman experiments were carried out with an in-house confocal spectrometer. The excitation source is a continuous Ar$^+$ laser  emitting at \unit{514.5}{\nano\metre} focused on the sample with a Mitutoyo 20$\times$ objective
to a spot of about \unit{2.4}{\micron}. The backscattered light was spatially filtered by a confocal pinhole, chromatically filtered by razor edge filters (Semrock), and then dispersed by a HR460 spectrograph (Horiba scientific) onto a CCD camera (Andor).

Angular dispersive x-ray diffraction experiments (XRD) were done at the PSICHE beamline  of the SOLEIL synchrotron facility. The x-ray wavelength was 0.3738 \AA. The focused x-ray beam had a spot size of \unit{10}{\micron} FWHM. Diffraction images were recorded on 2D detectors (Dectris Pilatus 2M CdTe or MarResearch marCCD). The detector distance and tilt, and position of the x-ray beam were calibrated using the x-ray standard  CeO$_2$. The integration of the x-ray images was performed with the Dioptas software~\cite{Prescher2015}. Le Bail or Rietveld refinements of the diffraction patterns were performed with the Fullprof software~\cite{carvajal_1993}.

\section{Results}

\subsection{Stability of AHH-II and II $\rightarrow$ DIMA transition at room temperature\label{sec:RTcompression}}
\begin{figure*}
\includegraphics[width=0.9\linewidth]{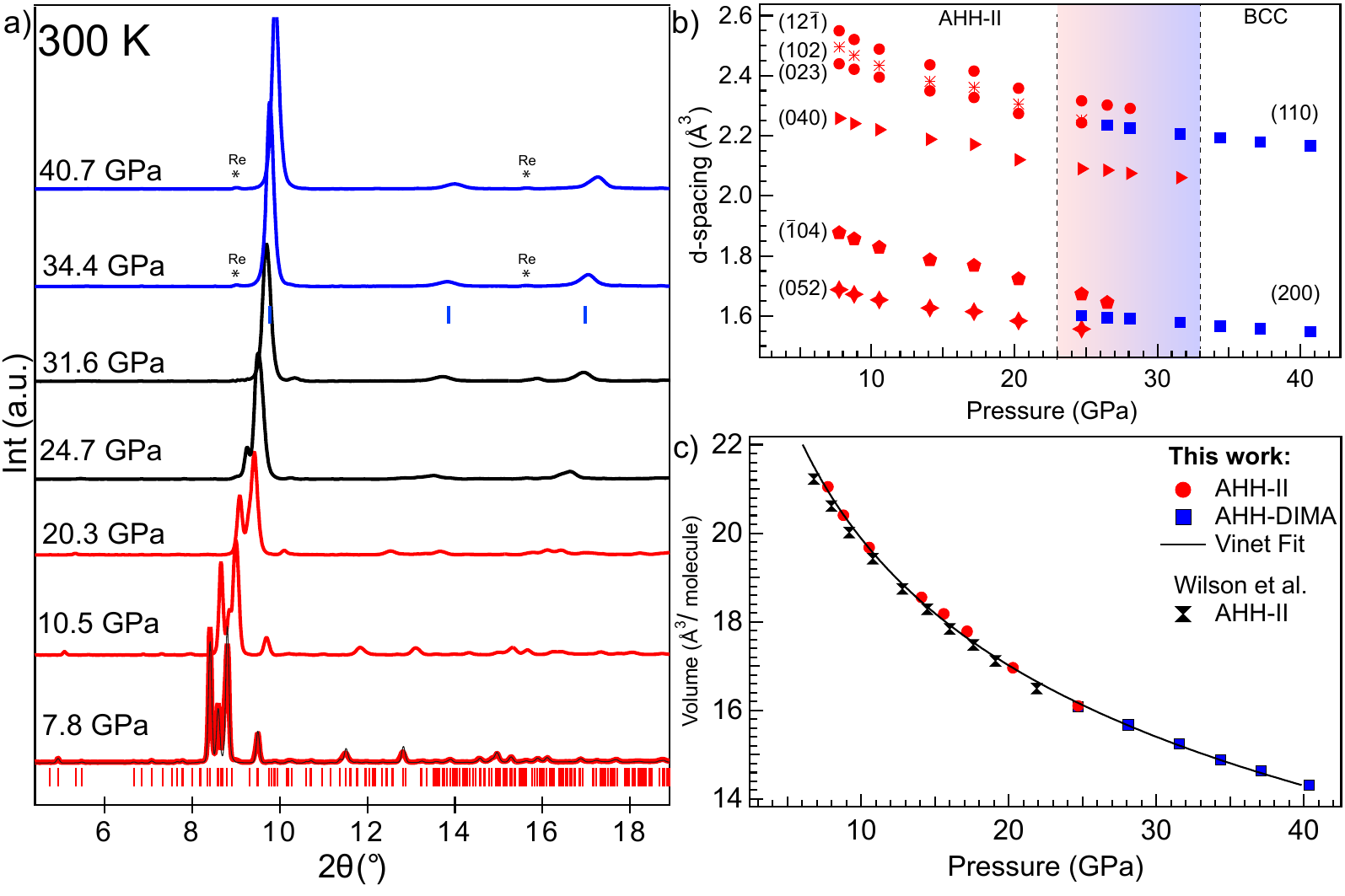}
\caption{\label{fig:DRX_compression_300K} (a) Powder XRD patterns of a AHH sample at 296~K on compression, showing the AHH-II (red lines) $\rightarrow$ AHH-DIMA (blue lines) transition. The patterns in black line correspond to the transition region. The Rietveld fit of the AHH-II pattern at 7.8 GPa is also shown as a black line. The blue and
red ticks show  the expected Bragg peak positions for AHH-DIMA and AHH-II, respectively. ''$\ast$'' indicates Bragg peaks from the Re gasket. Panel (b) shows the evolution with pressure of the interreticular distances corresponding to the most intense reflections of AHH-II and DIMA.  (c) Volume per molecule of AHH as a function of pressure at 296~K. The squares and circles correspond to experimental data for phase II and DIMA, respectively. The black line is a fit to all our data points using a Rydberg-Vinet equation of state. Black symbols are neutron diffraction data from Ref. \cite{Wilson2015, WilsonPhD2013}.}
\end{figure*}

In the first part of this work, we used XRD to study the stability and structural evolution with pressure of the AHH-II phase at room temperature. The AHH sample was initially at 7.8 GPa after loading. Its XRD pattern is consistent with the P$2_1/c$ structure of the AHH-II phase \cite{Wilson2012}, as shows the Rietveld refinement in Fig. \ref{fig:DRX_compression_300K}. No other phase was detected, hence the sample was pure AHH-II. The evolution of the XRD pattern as a function of pressure is shown in Fig.~\ref{fig:DRX_compression_300K}(a). From  7.9 to 22.2 GPa, the patterns remain consistent with the AHH-II phase. From 24.7 to 34.4 GPa, we observed a gradual transition to the DIMA I\emph{m$\bar{3}$m} phase: the peaks assigned to AHH-DIMA  progressively increased in intensity while those of AHH-II decreased. The sample fully transformed at 34.4 GPa. The evolution with pressure of the $d$-spacings of the first strong reflections are shown in  Fig.~\ref{fig:DRX_compression_300K}(b): as pressure increased, the AHH-II (102) and (023) reflections  coalesced to form the DIMA (110) peak while the (12$\overline{1}$) peak of AHH-II decreases in intensity until it disappears. 

The present results agree well with those of \citet{Wilson2015} obtained on a deuterated AHH sample using XRD. These authors located the II-DIMA transition between 26.6 and 34.8 GPa in the absence of intermediate pressure points. We show here that it occurs progressively in this pressure range. By contrast, we do not observe the sequence of two abrupt phase transitions at 19 and 26 GPa reported by \citet{ma2012}.

Fig. \ref{fig:DRX_compression_300K}(c) shows the measured volume per molecule of the AHH sample as a function of pressure in the range [7.8-40 GPa].  The present data agree very well with the neutron diffraction data on a deuterated AHH-II sample up to 26.6 GPa reported in Refs.\cite{Wilson2015, WilsonPhD2013}. As can be seen, the volume continuously decreases with pressure, and no discontinuous shift at the II-DIMA transition is observed within experimental uncertainties, estimated as 0.1 \AA$^3$/molecule. The complete data set is actually well fitted by a single Rydberg-Vinet equation of state \cite{Vinet1986}, giving a zero-pressure volume of 30.5(2) \AA$^3$/molecule, a zero-pressure bulk modulus $B_0$ of 6.7(3)~GPa, and a bulk modulus first pressure derivative $B'_0$ of 6.0(5). These values of $B_0$ and $B'_0$ are typical of H-bonded ices and within the range of previously reported values for pure \HHO \cite{Klotz2017} and  \NH \cite{Datchi_2006}.

\subsection{Stability of AHH-II and II--\pbcc transition at high temperature}

\begin{figure}
\centering
\includegraphics[width=1\linewidth]{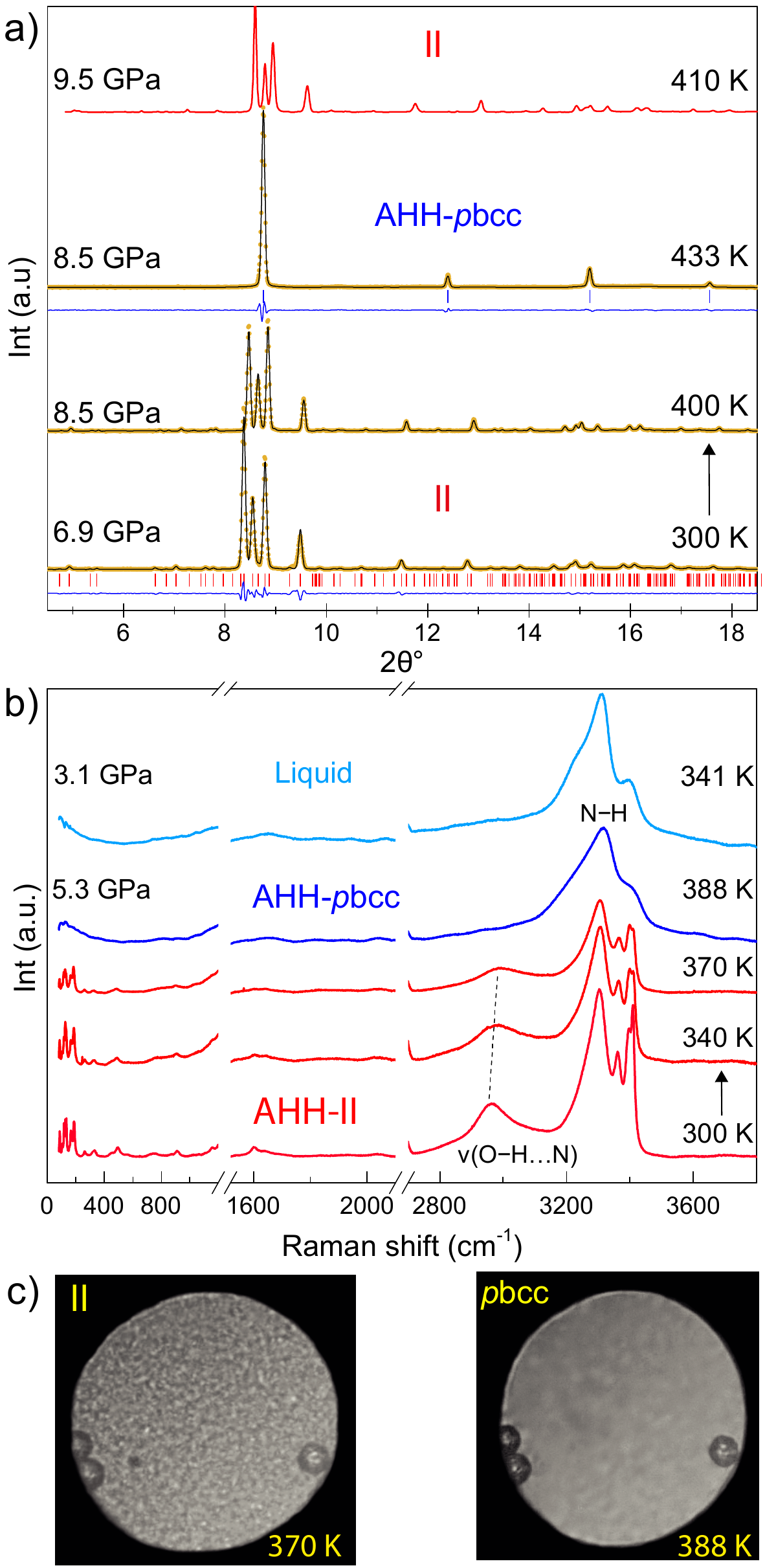}
\caption{\label{fig:transition_II-III_DRX_raman}(a) Evolution of the integrated XRD patterns of AHH obtained during the heating of phase II and showing the II-\pbcc transition. The black curves represent the results of the Rietveld refinements. The blue and red vertical lines respectively represent the expected Bragg reflections of cubic AHH-\pbcc and monoclinic AHH-II. The top pattern in red was obtained on cooling back the sample to 410 K, showing that it back-transited to phase II. (b) Evolution of Raman spectra of AHH as a function of T along an isobar at 5.3 GPa. The spectra for AHH-II and AHH-\pbcc are plotted in red and blue, respectively. The light blue line corresponds to the liquid phase. (c) Photographs at 5.3 GPa of (left) AHH-II at 370 K and (right) AHH-\pbcc at 388 K illustrate the disappearance of the birefringence at the II-\pbcc phase transition.}
\end{figure}

In a second step, we studied the stability of AHH-II at temperatures above 300 K using XRD and Raman scattering experiments. Fig.~\ref{fig:transition_II-III_DRX_raman}(a) presents the evolution of the powder XRD patterns as a function of temperature starting from a sample of AHH-II at 300 K and 6.9 GPa. AHH-II is observed up to 400 K, as confirmed by the Rietveld refinements shown in Fig.~\ref{fig:transition_II-III_DRX_raman}(a) (see also Fig. S1 and Tab. S1 of the SM~\cite{SM} for the atomic positions).  A  structural phase transition occurs between 400 K and 433 K at 8.5 GPa: the pattern at 433 K is  much simpler than that of AHH-II, consisting of 4 peaks which can be indexed with a \bcc unit cell with $a=3.463(3)$ \AA\ (see also the image plate in Fig. S2 of the SM~\cite{SM}). In the following, we will refer to this high temperature phase as AHH-\pbcc.  As seen in Fig.~\ref{fig:transition_II-III_DRX_raman}(a), AHH-II is recovered on cooling AHH-\pbcc at 410 K and 9.5 GPa. The II-\pbcc transition is thus reversible with small hysteresis. No coexistence between the two phases has been observed in an interval of $\sim 20$ K, indicating fast transition kinetics. We also note that the II-\pbcc transition results in a visual change of the sample: the reticulations observed in the birefringent phase II disappear in AHH-\pbcc which is a transparent and homogeneous solid, as illustrated in Fig.~\ref{fig:transition_II-III_DRX_raman}(c)

The refined lattice parameters of AHH-II and AHH-\pbcc are gathered in Tab. S2 of the SM~\cite{SM}. A volume difference of 0.6\% is calculated between phase II at 400 K and \pbcc at 433 K. The volume thermal expansion at 8.5 GPa of AHH-II is estimated from present data to be 1.29 (6) $\times 10^{- 5}$~K $^{-1}$, which gives a volume expansion of 0.047(8)\% for a temperature variation of 33 K. The observed volume jump of 0.6 \% between 400 K and 433 K is thus mainly due to the II-\pbcc phase transition, and characterizes the latter as a first-order transition.

As shown in Fig.~S1(b) of the SM~\cite{SM}, the XRD patterns of AHH-\pbcc can be refined using the same structural model as used for the DIMA phase~\cite{LovedayPRL_1999,Liu2017} and the plastic bcc phase recently reported in AMH \cite{Zhang2023}. The latter comprises of a bcc unit cell with O and N randomly distributed on site (0,0,0), with an occupation factor respecting the 1:2 ratio. No H atoms were included in the model since they are poor x-ray scatterers and strongly disordered in the DIMA and plastic phases.

The II-\pbcc phase transition was further studied by Raman spectroscopy in temperature ramps along two isobars at 5.3 and 9 GPa. The evolution of the Raman spectrum at 5.3 GPa as a function of T is shown in Fig.~\ref{fig:transition_II-III_DRX_raman}(b) (see also Fig. S3 of the SM~\cite{SM} for the evolution at 9 GPa). The Raman spectrum  abruptly changes between 370 and 388 K at 5.3 GPa as a result of the transition. The sharp lattice peaks of the ordered phase II disappear and only a featureless signal rising at low frequency is observed in AHH-\pbcc. Moreover, the band assigned to the $\nu$(O-H...N) stretching around 2900 \invcm in AHH-II \cite{Bertie_JCP1984} seems to disappear in AHH-\pbcc, suggesting that either H-bonds are absent or have a very short lifetime in the high-T phase. AHH-\pbcc is also characterized by broad N-H stretching modes in the region 3100-3400 \invcm, typical of hydrogen-disordered solids.

All the measured II-\pbcc transition points are plotted in Fig.~\ref{fig:AHH_melting}. As throughout this manuscript, the transition pressure and temperature values were taken as the mid-points between the last and first points of observation of the two phases, and the error bars are taken as the distance between the two points.  The transition line so obtained up to 10 GPa is well-fitted by the equation: 
\begin{equation}\label{eq1}
  T=T_0\left[ 1+(P-P_0)/a\right]^{(1/b)}  
\end{equation}
with $a$ = 1.87(4) GPa, $b$ =7.711(10), $P_0$ = 4.33 GPa and $T_0$ = 359 K are fixed and correspond to the lowest (P, T) conditions where  the II--\pbcc transition was observed.

As mentioned in the Introduction, the transition from AHH-II to a bcc crystal at high T has previously been reported by \citet{Wilson2015} from XRD experiments. \citet{Wilson2015} postulated that this phase was identical to the DIMA phase observed on compression at 300~K. We will however see below that there are clear and abrupt changes in the XRD and Raman patterns of the two phases that show they are different solids. The II--\pbcc transition points of Ref. \cite{Wilson2015} (see Fig.~\ref{fig:AHH_melting}) follow a similar trend as observed here, yet they are systematically about 15 K lower in temperature than ours, which may be explained by the fact that in Ref. \cite{Wilson2015}, the thermocouple was located further away from the sample compared to present experiments.

\subsection{Melting line of AHH}

\begin{figure}
	\centering
	\includegraphics[width=1\linewidth]{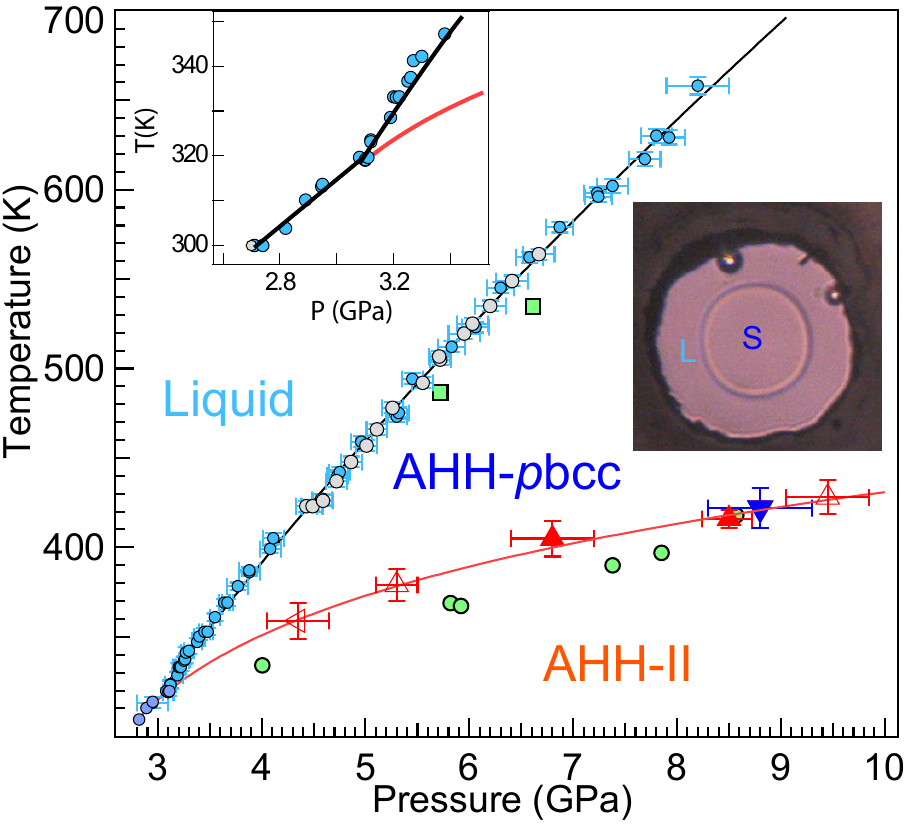}
	\caption{\label{fig:AHH_melting}AHH phase diagram for P$<$10 GPa. The solid and empty triangles correspond to the phase II-\pbcc transition observed by XRD and Raman spectroscopy, respectively. The orientation of the triangles indicates the thermodynamic path followed, and the P-T error bars correspond to the difference between the first observation of \pbcc and the last one of phase II, or \emph{vice-versa}. The grey and blue circles represent the melting points determined during this study in two different runs. The solid red and black lines represent fits of the II-\pbcc transition curve and of the melting curve, respectively, using the Simon-Glatzel equation \cite{Simon1929} (Eq. \ref{eq1}) The green circles and squares correspond, respectively,  to the II-\pbcc  and \pbcc-liquid transition points reported in Ref.~\cite{Wilson2015}. The inset shows a photograph of the solid AHH-\pbcc/fluid equilibrium at 630 K and 7.8 GPa. Two ruby balls are visible.}
\end{figure}

The melting properties of the AHH solid have been barely addressed in the literature: there are to date only two high-pressure melting points reported for AHH \cite{Wilson2015}, shown in Fig.~\ref{fig:AHH_Diagram_Intro}. Here the melting line has been determined on a fine grid of P-T points spanning the temperature range from 298 K to 650 K.

Melting of AHH was detected by visual observation of the solid-fluid equilibrium, as previously done for \HHO \cite{Datchi2000}, \NH \cite{Ninet2008}, and AMH \cite{Zhang2020}. The visual contrast between the two phases remained large enough to clearly distinguish the equilibrium up to 650 K (see inset of Fig.~\ref{fig:AHH_melting}). The measurement of pressure and  temperature at the coexistence defines a point $ (P_m, T_m) $ of the melting curve. As seen in Fig \ref{fig:transition_II-III_DRX_raman}(b), the Raman spectrum of AHH-\pbcc is very similar to that of the fluid phase, and it is thus difficult to infer melting of this phase solely from the Raman measurements. A similar behavior has been reported by \citet{Zhang2020} for phase VII of AMH.

\begin{comment}

\begin{table}[b]  
\caption{\label{Tabel:para_SG} Fit parameters of the Simon-Glatzel equation \\ $T_m=T\left[ (P_m-P)/a+1\right] ^{(1/b)}$}
\begin{ruledtabular}
\begin{tabular}{ccccc}
 & $T$ (K) & $P$ (GPa) & $a$ & $b$ \\
 \hline
P $\leq$ 3 GPa &  298(F)  & 2.71(F)  & 1.502(1) &  3.214(1)   \\
P $\geq$ 3  GPa &  320(F)  &  3.02(F) &  2.825(1)  &   1.557(3)  
\end{tabular}
\end{ruledtabular}
\end{table}
\end{comment}

\bgroup
\def\arraystretch{1.2}%  1 is the default, change whatever you need
\begin{table}
\caption{\label{Tabel:melting} Experimental melting points of AHH measured in this work.  $ P_m $ is in GPa and $ T_m $ is in K.}
\begin{ruledtabular}
\begin{tabular}{cccccc}
$P_m$	&	$T_m$	&	$P_m$	&	$T_m$	  &	$P_m$	&	$T_m$	\\
\hline	
2.70	&	300	&	3.55	&	361	&	5.32	&	475	\\
2.71	&	300	&	3.64	&	369	&	5.45	&	494	\\
2.82	&	304	&	3.67	&	369	&	5.55	&	492	\\
2.89	&	310	&	3.77	&	378	&	5.71	&	507	\\
2.95	&	314	&	3.88	&	386	&	5.72	&	505	\\
3.02	&	320	&	3.88	&	387	&	5.83	&	512	\\
3.10	&	319	&	4.08	&	399	&	5.95	&	519	\\
3.11	&	320	&	4.11	&	405	&	6.03	&	525	\\
3.12	&	325	&	4.43	&	423	&	6.06	&	523	\\
3.19	&	329	&	4.48	&	423	&	6.20	&	535	\\
3.21	&	333	&	4.49	&	423	&	6.30	&	545	\\
3.22	&	333	&	4.58	&	427	&	6.41	&	549	\\
3.24	&	338	&	4.59	&	426	&	6.58	&	562	\\
3.25	&	336	&	4.72	&	437	&	6.67	&	564	\\
3.26	&	337	&	4.73	&	440	&	6.87	&	579	\\
3.27	&	341	&	4.75	&	442	&	7.23	&	596	\\
3.29	&	342	&	4.86	&	447	&	7.30	&	598	\\
3.38	&	347	&	4.96	&	459	&	7.38	&	602	\\
3.40	&	350	&	5.01	&	457	&	7.70	&	617	\\
3.45	&	353	&	5.11	&	466	&	7.80	&	630	\\
3.48	&	353	&	5.26	&	478	&	7.90	&	629	\\
3.55	&	361	&	5.30	&	473	&	8.35	&	658	\\									
\end{tabular}
\end{ruledtabular}
\end{table}
\egroup

Two different samples were used to determine the melting curve of AHH in the range of 298-658 K and 2.7-8.2 GPa. The experimental melting points are plotted in Fig.~\ref{fig:AHH_melting} and reported in Table.~\ref{Tabel:melting}. The two measurements agree very well in the overlapping region (423-565 K).  A change of slope of the melting curve is observed at 3.0(1) GPa and 319(1) K due to the II-\pbcc phase transition. Indeed, we found that phase II congruently melts at 313 K, while at 325 K, we observed the congruent melting of the \pbcc phase (see the measured Raman spectra and the photographs of the sample during decompression and compression runs at 313 and 325 K in Fig. S4-S5 of the SM\cite{SM}). The observed change of slope thus highlights the triple point between II, \pbcc and the liquid phase at 3.0(1) GPa and 319(1) K. The melting points were fit separately below and above 3.0 GPa. For $P<3$ GPa, a linear regression adequately fits our 5 melting points of AHH-II with a slope of 52.3(1)~K.GPa$^{-1}$. For $P\geq 3$ GPa, Eq. \ref{eq1} gives an excellent fit of the AHH-\pbcc melting line, with $a=2.077(1)$~GPa, $b=1.768(1)$, and the fixed values $T_0=319$~K, $P_0=3.07$~GPa.

%\newpage

\subsection{Occurence and stability domain of  a new phase: AHH-\qbcc}

\begin{figure*}
\includegraphics[width=1\linewidth]{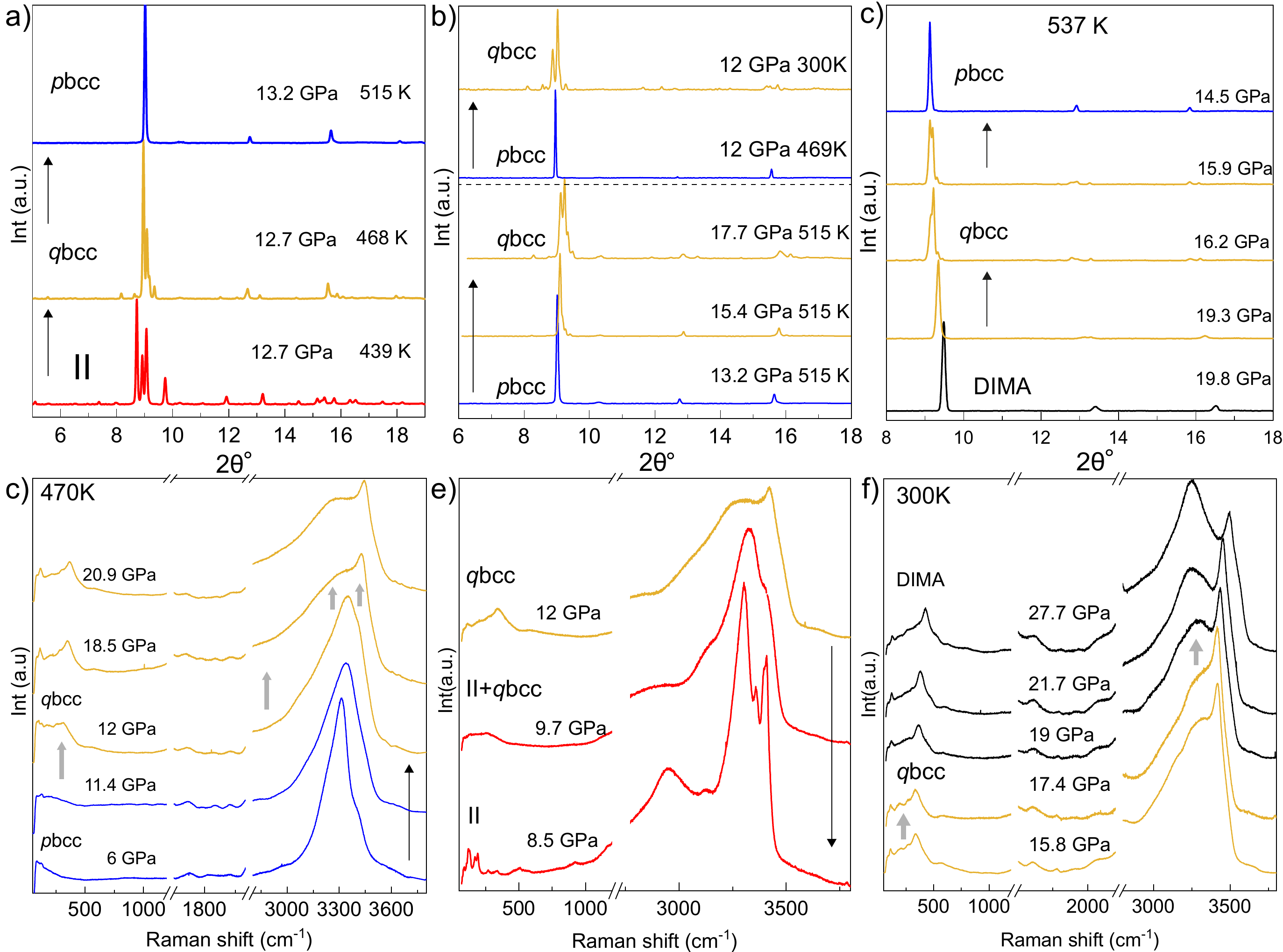}
\caption{\label{fig:Transition_II-IV-V_2} [a-c] XRD patterns of AHH collected along various paths crossing phase transitions: (a) isobaric heating at 12 GPa across the II-\qbcc and \qbcc-\pbcc transition; (b) the \pbcc-\qbcc transition is observed along an isothermal compression from 13.2 GPa to 20 GPa at 515 K and along an isobaric cooling at 12 GPa  (c) isothermal decompression from 19.8 GPa to 14.5 GPa at 537 K across the DIMA-\qbcc and \qbcc-\pbcc transitions. [d-f] Evolution of the Raman spectra following (c) a compression from 13.2 GPa to 20 GPa across the \pbcc--\qbcc transition at 470 K, (e) during the decompression from 12 GPa to 8.5 GPa at room temperature, and (f) during the compression at 300 K across the \qbcc--DIMA transition. The frequency windows 1200--1400 \invcm and 2200--2700 \invcm are omitted as they are dominated by, respectively, the first and second-order Raman signal from the diamond anvils. In all the graphs, the phases II, DIMA, \pbcc and \qbcc patterns or spectra are plotted in red, black, blue, and beige, respectively. The black arrows show the direction in which P or T varies in each experiment. }
\end{figure*}

Our investigations of AHH at pressures above 10 GPa and high temperature led to the discovery of another phase, AHH-\qbcc, whose P-T domain of stability is bordered by AHH-II, AHH-\pbcc on the low-pressure side, and DIMA on the high-pressure side, as described below.

Fig~\ref{fig:Transition_II-IV-V_2}(a) shows the evolution of the measured XRD patterns as a function of temperature on heating AHH-II along an isobar at 12.7 GPa. At 468 K, the pattern changes and cannot be indexed either by phase II, \pbcc, or a coexistence of the two phases. It thus corresponds to another phase of AHH, which we called AHH-\qbcc. When this solid is heated above 512 K at 13 GPa, the pattern changes again, and the observed peaks can be indexed with AHH-\pbcc (see Fig. 6 of the SM~\cite{SM} for the XRD image at the transition). As illustrated in Fig~\ref{fig:Transition_II-IV-V_2}(b), we also found that AHH-\qbcc can be obtained by isothermal compression at 515 K or by isobaric cooling at 12 GPa of the \pbcc phase. The transition line between these two phases was mapped up to $\sim$20 GPa, and the data points are plotted in Fig~\ref{fig:AHH_Driagram_V}.

Fig~\ref{fig:Transition_II-IV-V_2}(d) shows the evolution of the Raman spectrum with pressure along an isothermal compression of AHH-\pbcc at 470 K from 6 to 20 GPa. The \pbcc-\qbcc transition occurs in-between 11.4 and 12 GPa, signalled by several changes in the Raman spectrum indicated by gray arrows in Fig~\ref{fig:Transition_II-IV-V_2}(d): (1) the appearance of lattice modes below 700 \invcm in \qbcc which are absent in \pbcc, (2) an increase in the Raman signal at 2800-3000 \invcm, a region assigned to the O-H...N stretching modes in AHH-II, and (3) a broadening of the N-H stretching band and the appearance of a mode at $\sim$ 3500 \invcm.

Upon increasing pressure, we found that the \qbcc phase transits into the DIMA phase at all investigated temperatures. This transition was studied in the range 500-600 K by XRD, where it occurs around 19 GPa in both compression and decompression runs [see Fig~\ref{fig:Transition_II-IV-V_2}(c)], showing that the transition is reversible. Using Raman spectroscopy, we followed the compression of  a \qbcc sample at 300 K from 15.8 GPa to 27.7 GPa. The evolution of the Raman spectra is shown in Fig~\ref{fig:Transition_II-IV-V_2}(f), and exhibit clear changes  in the spectra collected below and above 19 GPa, highlighted by arrows in the figure. First, there is a sharp increase in the intensity of the N-H stretching modes at $\sim$3300 \invcm. Second,
the two lattice peaks at  190 and 290, visible on the spectrum at 15.8 and 17.4 GPa, disappear at 19 GPa . The \qbcc-DIMA transition point at 18.2(8) GPa at 300 K thus defined lines up well with the transition points obtained by XRD at high temperature. 

All the transition points between AHH-II, \pbcc, DIMA and \qbcc are gathered on the phase diagram of Fig.~\ref{fig:AHH_Driagram_V}. So far the structure of \qbcc remains undetermined (see section \ref{sec:qbcc}), thus we cannot assert whether these are first-order transitions, however the three transitions were observed to be sharp. The \qbcc--\pbcc and \qbcc-DIMA transitions are reversible, but this is not the case for the II--\qbcc transition, as shown in Fig~\ref{fig:Transition_II-IV-V_2}(b): once \qbcc is formed at high temperature above 10 GPa, it is kept down to room temperature. Phase II was only recovered when \qbcc was decompressed below 9.7 GPa at 300 K as shown in Fig~\ref{fig:Transition_II-IV-V_2}(e). We did not investigate in detail the II--\qbcc transition line above 300 K but noted that on extrapolating the fit of the \pbcc--\qbcc line, the latter passes through the measured II--\qbcc transition point at 300 K (see Fig. \ref{fig:AHH_Driagram_V}). Our observations thus indicate that \qbcc is more stable than II above 10 GPa and for T$\geq$300 K. As seen in section \ref{sec:RTcompression} though, the \qbcc phase is not observed when compressing phase II at 300 K, the latter transiting directly to the DIMA phase above 25 GPa. This suggests that the kinetic barrier between phases II and \qbcc is too large at 300 K to observe the II--\qbcc transition upon compression, making it necessary to heat phase II to obtain phase \qbcc.
Finally, it can be seen that the transition lines \pbcc--\qbcc and \qbcc--DIMA meet at high temperature and delimit a domain of stability of phase \qbcc which closes at high temperature. The intersection of the two lines defines a triple-point DIMA--\pbcc--\qbcc located at 21.5 GPa and 630 K.

\subsection{Isostructural \pbcc $\rightarrow$ DIMA transition at high temperature}

\begin{figure*}
\includegraphics[width=1\linewidth]{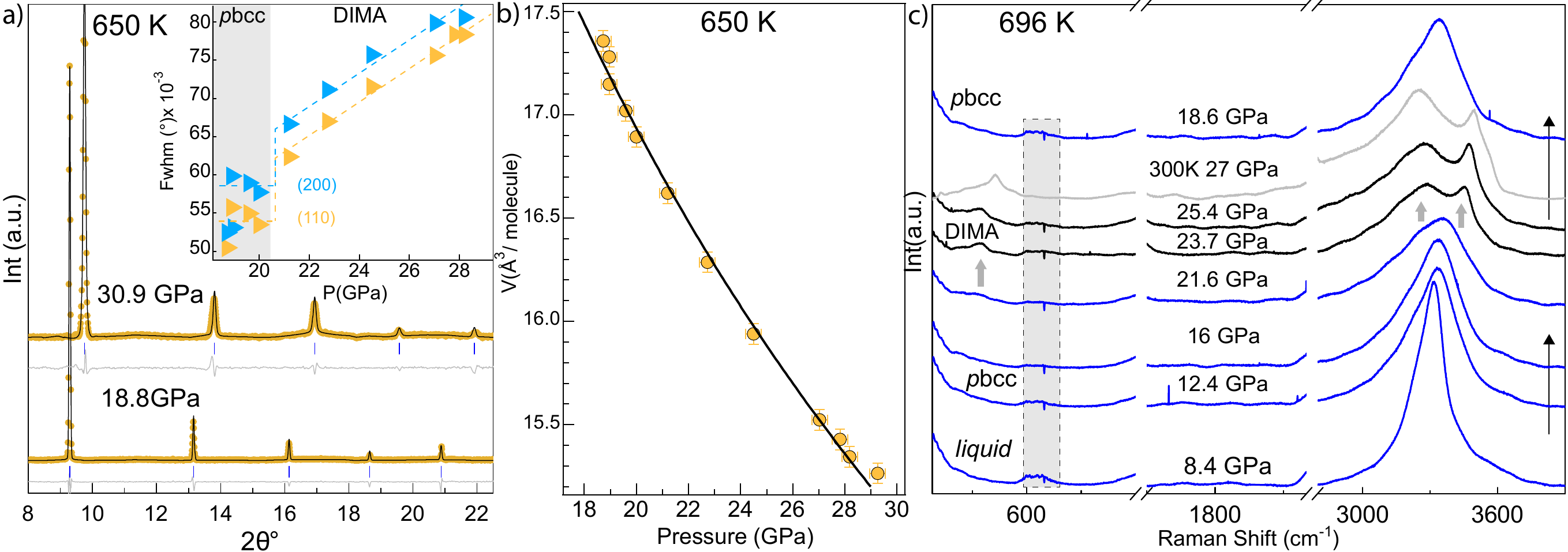}
\caption{\label{650_isotherme} (a) XRD patterns measured at the lowest pressure (18.8 GPa) and the highest pressure (30.9 GPa) during an isothermal compression at 650 K. The black curves are Le Bail refinements and ticks show the positions of the Bragg peaks. The inset shows the full width at half-maximum (FWHM) evolution of the first two Bragg peaks as function of pressure. (b) Volume per molecule as a function of pressure at 696 K. (c) Evolution of the Raman spectra along the isothermal compression at 696 K from 8.4 to 25.4 GPa and then decompression to 18.6 GPa, through the \pbcc$\leftrightarrows$DIMA transition. The blue and black spectra correspond respectively to phase \pbcc and DIMA. The gray spectrum corresponds to phase DIMA obtained at 27 GPa and 300 K, to compare with the spectrum obtained at 696 K (black).}
\end{figure*}

As seen above, one expects a direct transition between the \pbcc and the DIMA phase above the \pbcc-\qbcc-DIMA triple point. To investigate this, we performed Raman measurements upon isothermal compressions of phase \pbcc at 660 K and 696 K.  As seen in Fig~\ref{650_isotherme}(c) (see also Fig. S8 of the SM~\cite{SM} for the data at 660~K), the Raman peaks of \pbcc were observed from 12.4 to 21.6 GPa at 696 K.  At 23.7 GPa, the Raman spectrum changed, notably with the appearance of a peak at 3440 \invcm and an increase in Raman intensity around 3200 \invcm. This spectrum can be compared with that of the DIMA phase  at 27.7 GPa  and 300 K in Fig~~\ref{650_isotherme}(c) (in black and grey color, respectively): their similarity strongly suggests that \pbcc transits to DIMA in-between 21.6 and 23.7 GPa at 696 K. On decompression, the Raman spectrum of \pbcc is recovered at 18.6~GPa, which shows the reversible nature of the \pbcc-DIMA transition (no Raman spectra was collected on decompression from 25.4 to 18.6 GPa, so the back-transition pressure is ill-defined).

We also collected XRD patterns along an isothermal compression at 650 K starting from \pbcc at 18.7 GPa and up to 30.9 GPa. The patterns recorded at the start and the end of the compression are presented in Fig~\ref{650_isotherme}(a). All patterns can be indexed by a bcc unit cell, and the \qbcc phase was not observed, confirming that its P-T domain closes at high temperature.  The evolution with  pressure of the volume per molecule at 650 K is shown in Fig. \ref{650_isotherme}(b). A small volume discontinuity can be seen at 19 GPa but its magnitude remains within the uncertainty of the measurement, and a single Vinet equation of state adequately fits the full volume data set. As visible in Fig~\ref{650_isotherme}(a), a notable difference between the XRD patterns at 18.7 GPa and 30.9 GPa resides in the width of the diffraction peaks. In the insert, we plotted the full width at half maximum (FWHM) of the first two diffraction peaks, (110) and (200), as a function of pressure. These are roughly constant up to 21 GPa where a sudden increase occurs. At higher pressure, the peak widths continue to increase but in a gradual fashion. This correlates well with the changes seen in the Raman spectrum, suggesting that the sudden broadening of XRD peaks is a signature of the \qbcc-DIMA transition. 

\begin{figure}
\centering
\includegraphics[width=1\linewidth]{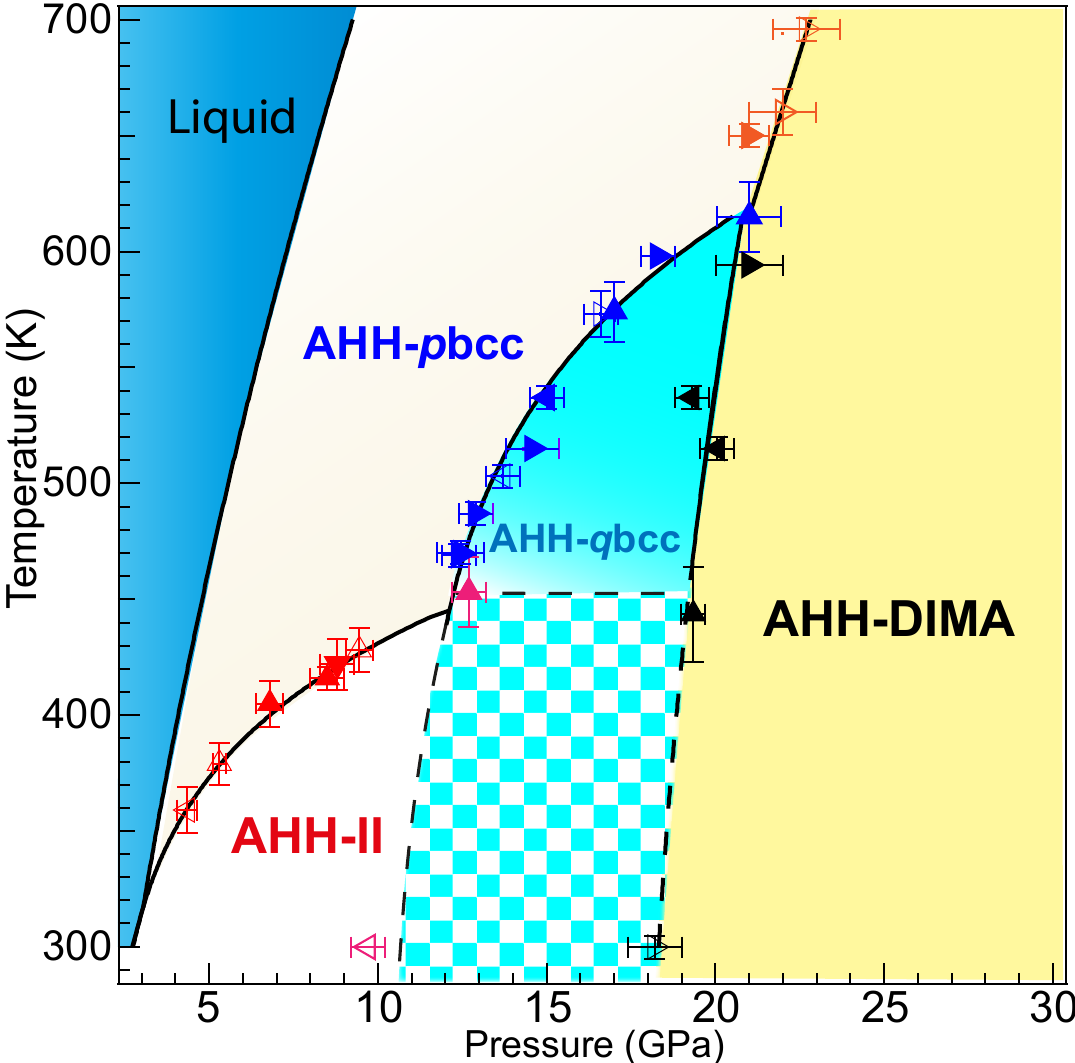}
\caption{\label{fig:AHH_Driagram_V}Experimental phase diagram of AHH P$<$30 GPa determined during this work. The symbols represent the transition points \pbcc--\qbcc (in blue), II--\qbcc (in red), DIMA--\qbcc (in black), \pbcc-DIMA (in orange) and II-\qbcc (in pink) observed by X-ray diffraction (solid triangles) and by Raman spectroscopy (open triangles). The orientation of the triangles indicates the followed thermodynamic path, and the P-T error bars correspond to the difference between the first and last points of the observed transition. The horizontal dotted line indicates the kinetic transition from II to \qbcc upon isobaric heating at 12 GPa. The blue-white checkerboard represents the P-T domain where phase II is observed on compression but is metastable.}
\end{figure}

All the measured \qbcc-DIMA transition points are shown in Fig~\ref{fig:AHH_Driagram_V}. The sharp changes observed in both Raman and XRD experiments suggest that this transition is first-order although no volume discontinuity could be detected within our experimental accuracy (the latter was estimated to be about 0.1 \AA$^3$/molecule based on the standard deviation between the volume calculated from individual peaks and the average one). Since the two phases can be described by the same space group and arrangement of O and N atoms, the transition is isosymmetric and the difference between the two phases must reside in the configuration of H atoms. As a matter of fact, we present below several arguments suggesting that \pbcc is a crystalline plastic form of the DIMA phase.

\section{Discussion}
\subsection{Plastic nature of AHH-\pbcc}

Plastic solids are molecular crystals where molecules exhibit a dynamic orientational disorder, i.e., they behave like free rotors whose center of mass remains bound to the lattice site. Well known examples of plastic solids are the ammonia phases II and III \cite{Doverspike1986}. Plastic forms of water \cite{takii2008, Hernandez2018} and ammonia hydrates \cite{Naden2020} have also been predicted by calculations. Recently, \citet{Zhang2023} have shown the occurrence of a plastic phase in ammonia monohydrate (AMH), AMH-VII, that is a plastic form of the DIMA phase.

\begin{figure}[]
\centering
\includegraphics[width=1\linewidth]{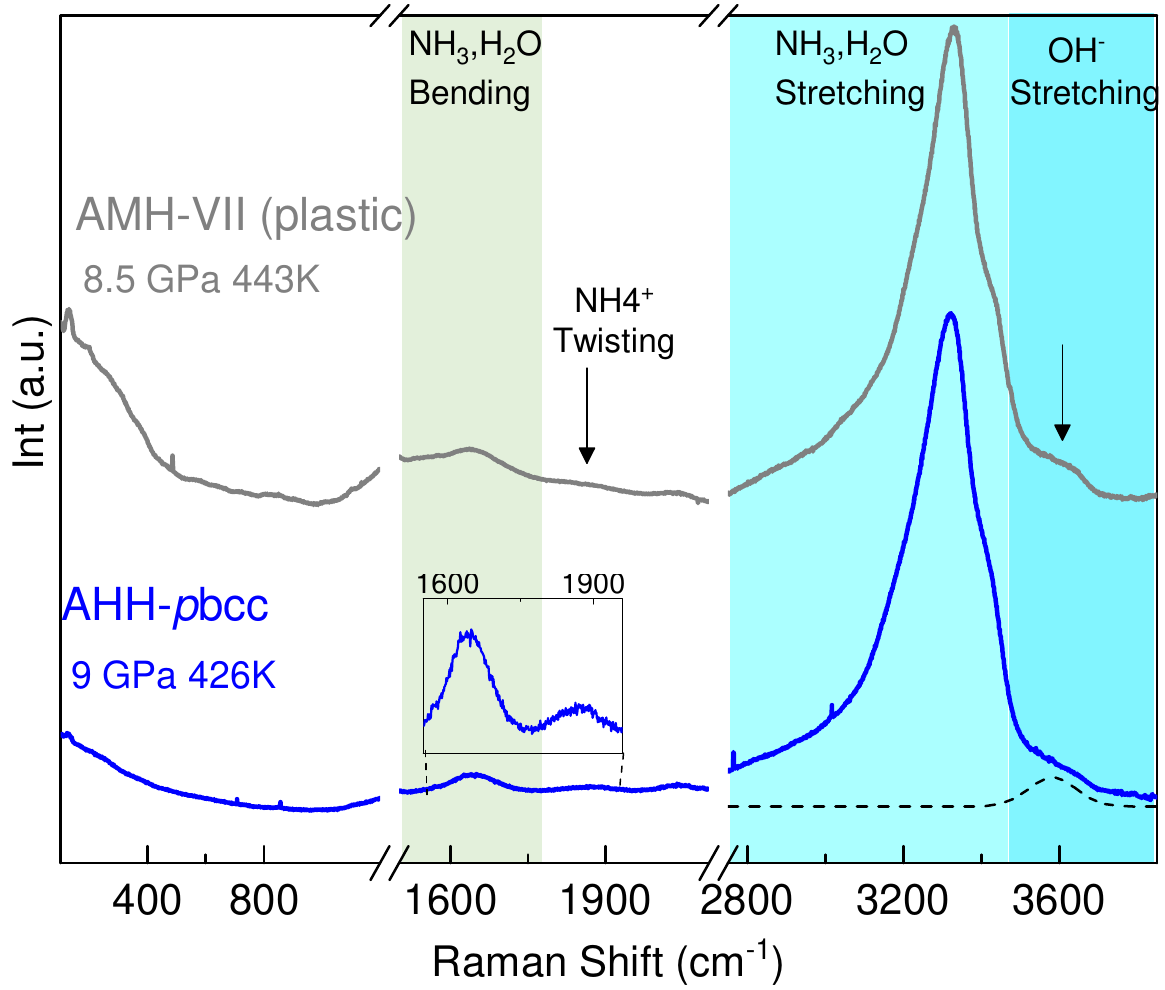}
\caption{\label{fig:Raman_AHH_AMH_plastic} Comparison between the measured Raman spectra of
AHH-\pbcc (blue), and AMH-VII (gray) at P-T conditions indicated in the figure. The frequency windows 1200-1400 \invcm and 2300-2700 \invcm dominated by, respectively, the first and second-order Raman signals from the diamond anvils have been omitted. The arrows emphasize bands assigned to ionic species in AMH-VII ~\cite{Zhang2023}, which are also present in the spectrum of AHH-\pbcc.}
\end{figure}

Several experimental observations made above indicate that phase \pbcc is a plastic solid. First, AHH-\pbcc has strong similarities with the AMH-VII phase : (1) they have the same bcc structure; (2) their Raman spectra are very similar as shown in Fig.~\ref{fig:Raman_AHH_AMH_plastic}. Furthermore the Raman spectrum of phase \pbcc is very similar to that of the liquid of the same composition, which has already been observed for the plastic and liquid phase of AMH and pure \NH \cite{Kume2001,Zhang2023}.

A second argument comes from the entropy increase at the transition between the proton-ordered phase II and \pbcc. The latter can be deduced from the measured volume jump $\Delta V$ and the slope of the transition line \emph{via} the Clausius-Clapeyron relation. At 8.5 GPa, we measured $\Delta V=3.94\times 10^{-3}$ cm$^3$/g and a slope of 0.108 GPa.K$^{- 1} $, which gives an entropy increase of 0.99(5)$R$ from II to \pbcc, where $R$ is the ideal gas constant.  This value is identical, within uncertainties, to that measured in pure ammonia at the transition between the ordered solid IV and the plastic phase III [1.04 (4)R] \cite{Ninet2012}. Moreover, \citet{Ninet2012} showed that for \NH this entropy variation corresponds to the expected configurational entropy difference between rotationally ordered and disordered configurations.

Thirdly, \pbcc is characterized by very sharp Bragg peaks compared to those of DIMA. The broadening of the Bragg peaks in the DIMA phase is due to the local distortion of the cubic lattice, induced by the different types of hydrogen bonding between ions and molecules~\cite{Liu2017}. The presence of sharp Bragg peaks in phase \pbcc indicates that such local deformation is absent. This could be understood if phase \pbcc presents a dynamic orientational disorder since in this case, the lifetime of the H bonds is very short, which is also consistent with the disappearance at the II-\pbcc transition of the O--H...N stretch Raman band.

It is also interesting to note that the \pbcc-DIMA transition is reminiscent of the behaviour of water ice at the plastic ice-- ice-VII transition as described by \citet{Hernandez2018}. Indeed, this transition in \HHO is isostructural and only involves a change in the dynamics of the hydrogen atoms, from a static rotational disorder to a dynamic one, without any change in the oxygen sub-lattice.

A direct confirmation of the plastic nature of AHH-\pbcc requires to perform experiments that directly probe the dynamics of hydrogen atoms. This could be performed using quasi-elastic neutron scattering, as recently done in Ref. \cite{Zhang2023} for AMH.

\subsection{Structural and vibrational characteristics of AHH-\qbcc\label{sec:qbcc}}

As seen above, the present study has uncovered the presence of a previously unknown phase of AHH, called AHH-\qbcc, stable in the intermediate P-T range between AHH-II, AHH-\pbcc and DIMA. We discuss below the structural and vibrational information presently available for this phase based on our XRD and Raman experiments.

The XRD patterns of AHH-\pbcc are presented in the panels (a-c) of Fig. \ref{fig:Transition_II-IV-V_2}. It can be noted a large variability in the relative intensities of the Bragg peaks, which comes from the fact that this phase was usually produced in polycristalline form with strong texture. A nice powder pattern could hovewer be obtained upon heating a powder sample of AHH-II at 12.7 GPa, shown in Fig. \ref{fig:Transition_II-IV-V_2}(a) (see Fig. S6 of the SM~\cite{SM} for the corresponding diffraction image).                                                  

\begin{figure}
	\centering
	\includegraphics[width=0.5\textwidth]{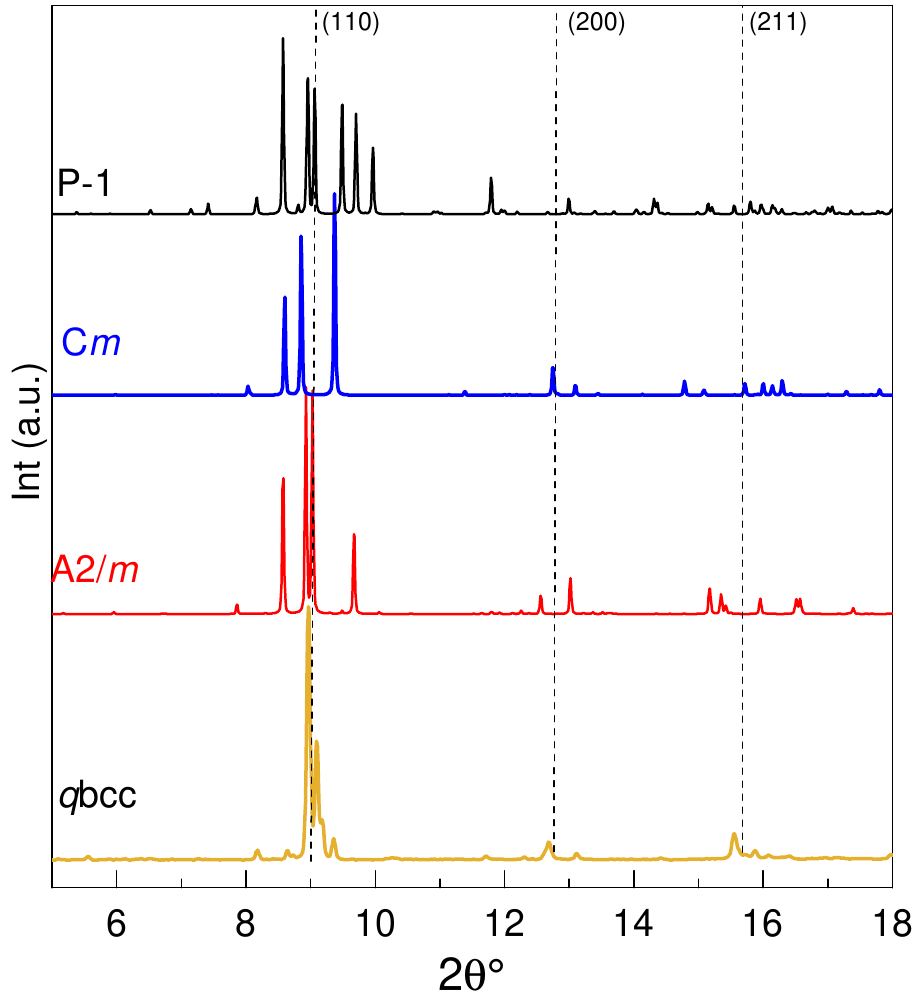}
\caption{\label{fig:DRX_qbcc_vs_predictio} Comparison between the experimental XRD pattern of AHH-\qbcc at 12.7 GPa and 468 K (yellow line) with the simulated patterns at 12 GPa and 0 K of the "quasi-bcc" A$2/m$ (red), C$m$ (blue), P$\overline{1}$ (black) phases predicted by Ref.~\cite{Naden2017}. The vertical dashed lines indicate the positions of the first three Bragg peaks in AHH-\pbcc at the same pressure.}
	\end{figure}

For the same angular range, the number of Bragg peaks from phase \qbcc is much larger than those from \pbcc, showing that the \qbcc is of lower symmetry. It can be noticed however that the diffracted intensity from \qbcc is concentrated around the same angular positions as those of the bcc reflections of phase \pbcc, signaling a relation between the two structures. In their theoretical work, \citet{Naden2017} predicted that AHH-II transforms at 23 GPa and 0 K into a partially ionic structure of space group A$2/m$, composed of \NH, \OHm and \NHP. In this fully ordered structure, O and N atoms are located on positions close to that of a bcc lattice, hence A$2/m$ was  denoted as ''quasi-bcc''. Ref. \cite{Naden2017} also found that several other quasi-bcc structures have slightly higher energies than A$2/m$ in the same pressure range. Fig. \ref{fig:DRX_qbcc_vs_predictio} compares the simulated XRD pattern at 12 GPa for these theoretical quasi-bcc crystals to the experimental one of AHH-\qbcc at 12.7 GPa, 468 K: none of them is compatible with experiment, thus AHH-\qbcc has a different structure than those predicted by Ref. \cite{Naden2017}.

Our attempts to solve the structure of the \qbcc phase have so far not been fully successful and this work is still in progress. The observed reflections are compatible with a monoclinic unit cell with lattice parameters $a = 5.4606(4)$  \AA , $b  = 7.8050(6)$  \AA , $c= 5.2158(4)$  \AA\  and $\beta = 91.78^{\circ}$, and most probable space group P$2/m$. A Le Bail refinement of the experimental pattern using this model is shown in Fig. S9 of the SM~\cite{SM}. This gives a unit cell volume of 222.21 (3) \AA$^3 $ at 12.7 GPa, which, based on the equation of state of AHH at 300 K, can accommodate a maximum of 12 molecules. Despite the good agreement with the experimental pattern, we note that the $P2/m$ structure has low-angle reflections which are not observed (see insert in Fig.~ S9), and at this stage, we are thus unable to ascertain whether this structure is correct.

The Raman spectra of the \qbcc phase are illustrated in Fig. \ref{fig:Transition_II-IV-V_2}(d-f). It can be noted that the Raman spectra of \qbcc and of the DIMA phase show strong similarities. The low-frequency region hosting the lattice modes is dominated in both phases by a broad band extending form 500 \invcm down to the lowest measured frequencies and peaked at $\sim350$ \invcm. In \qbcc howvever, this band presents two more resolvable humps around 230 and 270 \invcm. The band around 1600~\invcm assigned to $\nu_2(\HHO)$ and $\nu_4(\NH)$ have the same shape and intensity. The one at 2900 \invcm is assigned to the O-H stretching of H-bonded molecules or ions and is similar in the two phases. In the region of the N--H stretching ($\sim3300$ \invcm), both phases display a broad band and a sharper high-frequency peak, yet the two are further separated and thus better resolved in DIMA. We also note a shoulder peak at about 3600 \invcm which is more intense in \qbcc. This band could be assigned to O-H stretch modes from weakly bound or unbound molecules or ions. The strong resemblance of the Raman spectra of \qbcc and DIMA suggest that \qbcc is, like DIMA, composed of molecules (\HHO and \NH) and ions (\OHm and \NHP). Moreover, the presence of broad lattice and vibronic bands indicates that \qbcc is a proton-disordered phase. 

\section{Conclusions}

In this paper, we have presented a detailed and comprehensive study of the phase diagram of AHH in the P-T domain [0-30 GPa, 300-700 K], combining x-ray diffraction and Raman spectroscopy experiments, and visual observations. To our knowledge, this is the first complete investigation of the phase diagram of AHH in this P-T range. 4 solid phases, AHH-II, AHH-\pbcc, AHH-\qbcc and DIMA have been identified and the final phase diagram is drawn in Fig.~\ref{fig:AHH_Driagram_V}. 

We first studied the stability of AHH-II under comrpession at 300 K and confirmed the transition to the DIMA phase at pressures above 24 GPa. This transition occurs gradually over a 10 GPa range and completes at about 34 GPa without any apparent volume discontinuity.  We then mapped the transition line between phase II and \pbcc. The latter, previously observed by \citet{Wilson2015}, is stable at high temperature up to the melting line and exhibits a bcc structure with $Im\bar{3}m$ space group. We then measured the melting curve of AHH up to a maximum temperature of 658 K and observed that (1) it is congruent in this P-T domain and (2) it exhibits a triple point II--\pbcc--L at 3.0(1) GPa--319(1) K.

This exploration of the phase diagram allowed us to uncover the existence of a previously unknown phase noted AHH-\qbcc. This phase was obtained either by heating phase II above 450 K at 12 GPa, or by compressing phase \pbcc above 450 K. It is therefore only formed at high temperature but once formed, it does not transit back to phase II on cooling down to room temperature. Similarly, the decompression of phase DIMA at 300 K leads to phase \qbcc at about 19 GPa, which itself transits into phase II at 9.5 GPa. We have therefore concluded that phase \qbcc is the stable phase between $\sim$10 and 19 GPa for T$\leq$450 K, and that phases II and DIMA observed in this pressure range are metastable.  The study of the transition lines \pbcc-\qbcc and \qbcc-DIMA showed that the stability domain of AHH-\qbcc closes at high temperature with a DIMA-\pbcc-\qbcc triple point at 21.5 GPa and 630 K. Finally, we observed sharp changes in the XRD patterns and Raman spectra at the \pbcc-DIMA isostructural transition which shows that these two phases are distinct unlike previously thought.

We have presented several arguments that suggest that phase \pbcc is plastic. No sign of plasticity is observed in DIMA, suggesting that these two phases mainly differ in their proton dynamics. The same situation was previously detected in AMH, where the hydrogen dynamics were directly probed by quasi-elastic neutron scattering (QENS)\cite{Zhang2020}. It would thus be valuable to perform QENS experiments on AHH to confirm the plasticity of AHH-\pbcc. It would also be very interesting to extend the investigation of the phase diagram to higher P-T conditions and determine in particular whether AHH becomes a superionic solid as reported previously in both pure \NH \cite{Ninet2012} and pure \HHO \cite{Queyroux2020}. Such a superionic state has been predicted by calculations in the ammonia hydrates \cite{Bethkenhagen2015,Naden2018} but lacks experimental observation so far.

\begin{acknowledgments}
The authors thank J.P. Iti\'{e} for his assistance in the XRD experiments at the PSICHE beamline and L. Delbes for his assistance in the use of the XRD platform at IMPMC for the preliminary sample characterizations. We acknowledge financial support from the Agence Nationale de la Recherche under grant ANR-15-CE30-0008-01 (SUPER-ICES), the SOLEIL synchrotron facility for the provision of beam time to proposals 20170522 and 20180575, and the spectroscopy and XRD platforms of IMPMC.
\end{acknowledgments}

\bibliographystyle{apsrev4-1}
%\bibliography{references}

%

\end{document}